\documentclass[twocolumn,superscriptaddress,aps,preprintnumbers,amsmath,amssymb,sort&compress,nofootinbib
]{revtex4-1}
\usepackage{amsfonts,amsmath,amssymb,ascmac,bm,tensor}
\usepackage{comment}
\usepackage{ifpdf}
\usepackage{slashed}
\usepackage{color}
\usepackage[mathscr]{eucal}
\usepackage[utf8]{inputenc}

\usepackage{cancel}

\ifpdf
  \usepackage{graphicx}     %   usepackage without driver option
  \usepackage[bookmarksopen,colorlinks=true,linkcolor=bblue,citecolor=bblue,urlcolor=ppink]{hyperref}
\else     % For (p)LaTeX + dvipdfmx
  \usepackage[dvipdfmx]{graphicx}     %   usepackage with driver option
  \usepackage[dvipdfmx,bookmarksopen,colorlinks=true,linkcolor=bblue,citecolor=ppink,urlcolor=darkred]{hyperref}
\fi

\definecolor{red}{rgb}{1,0,0}
\definecolor{darkred}{rgb}{0.6,0,0}
\definecolor{darkgreen}{rgb}{0.992447,0.623778,0.034597}
\definecolor{ppink}{rgb}{1,0.4,0.4}
\definecolor{bblue}{rgb}{0.284602,0.317763,0.963947}
\definecolor{purple}{rgb}{0.5 ,0, 0.7}

\newcommand{\dd}{\mathrm{d}}
\newcommand{\R}{\text{R} }

\newcommand{\GeV}{\text{GeV} }
\newcommand{\tmax}{\text{max} }

\makeatletter
\newcommand\footnoteref[1]{\protected@xdef\@thefnmark{\ref{#1}}\@footnotemark}
\makeatother

\allowdisplaybreaks[1]

\begin{document}

%%%%%%%%%%%%%%%%%%%%%%%%%%%
%%%%%%%%%%% Title %%%%%%%%%%%
%%%%%%%%%%%%%%%%%%%%%%%%%%%

\title{
Enhancement of Gravitational Waves Induced by Scalar Perturbations \\ due to a Sudden Transition from an Early Matter Era to the Radiation Era
}
\author{Keisuke Inomata}
\affiliation{ICRR, University of Tokyo, Kashiwa, 277-8582, Japan}
\affiliation{Kavli IPMU (WPI), UTIAS, University of Tokyo, Kashiwa, 277-8583, Japan}
\author{Kazunori Kohri}
\affiliation{Kavli IPMU (WPI), UTIAS, University of Tokyo, Kashiwa, 277-8583, Japan}
\affiliation{Theory Center, IPNS, KEK, 1-1 Oho, Tsukuba, Ibaraki 305-0801, Japan}
\affiliation{The Graduate University for Advanced Studies (SOKENDAI), 1-1 Oho, Tsukuba, Ibaraki 305-0801, Japan}
\author{Tomohiro Nakama}
\affiliation{Institute for Advanced Study, The Hong Kong University of Science and Technology, \\Clear Water Bay, Kowloon, Hong Kong}
\author{Takahiro Terada}
\affiliation{Theory Center, IPNS, KEK, 1-1 Oho, Tsukuba, Ibaraki 305-0801, Japan}

\begin{abstract}
\noindent
We study gravitational waves induced from the primordial scalar
perturbations at second order around the reheating of the Universe.
We consider reheating scenarios in which a transition from an early
matter dominated era to the radiation dominated era completes within a
timescale much shorter than the Hubble time at that time. We find that
an enhanced production of induced gravitational waves occurs just
after the reheating transition because of fast oscillations of
scalar modes well inside the Hubble horizon.  This enhancement
mechanism just after an early matter-dominated era is much more
efficient than a previously known enhancement mechanism during an
early matter era, and we show that the induced gravitational
waves could be detectable by future observations if the reheating
temperature $T_{\text{R}}$ is in the range
$T_\R \lesssim 7\times 10^{-2}$GeV or
$20 \, \text{GeV} \lesssim T_\R \lesssim 2 \times 10^7 \, \text{GeV}$.
This is the case even if the scalar perturbations on small scales are
not enhanced relative to those on large scales, probed by the
observations of the cosmic microwave background.
\end{abstract}

\date{\today}
\maketitle
\preprint{IPMU 19-0067}
\preprint{KEK-TH-2122}
\preprint{KEK-Cosmo-237}

%%%%%%%%%%%%%%%%%%%%%%%%%%%%%%
\section{Introduction}
\label{sec:intro}
%%%%%%%%%%%%%%%%%%%%%%%%%%%%%%

Recently, gravitational waves (GWs) have been attracting more and more
attention.  So far, LIGO and Virgo
collaborations have succeeded in detecting GWs from merging black
holes~\cite{Abbott:2016blz,Abbott:2017oio}.  KAGRA is also expected to
detect GWs in a few years~\cite{Akutsu:2018axf}.  GWs provide a lot of
information about not only the nature and the origins of black holes,
but also about the early Universe.  Stochastic GWs induced by curvature
perturbations at second order are one of the cosmological GW sources
closely related to the study of the early
Universe~\cite{Ananda:2006af,Baumann:2007zm,Saito:2008jc,Saito:2009jt,Alabidi:2012ex}.
There are a number of recent studies about such induced
GWs~\cite{Espinosa:2018eve,Kohri:2018awv,Cai:2018dig,Bartolo:2018evs,Bartolo:2018rku,Unal:2018yaa,Byrnes:2018txb,Inomata:2018epa,Clesse:2018ogk,Cai:2019amo,Cai:2019jah,Wang:2019kaf,Ben-Dayan:2019gll,Tada:2019amh},
some of which are related to primordial black holes.  The induced GWs,
along with other GW backgrounds of astrophysical as well as
cosmological origins, can be investigated by the ongoing or future GW
projects, such as pulsar timing array observations
(EPTA~\cite{Lentati:2015qwp}, PPTA~\cite{Shannon:2015ect},
NANOGrav~\cite{Arzoumanian:2015liz},
SKA~\cite{Moore:2014lga,Janssen:2014dka}), ground based interferometer
experiments (advanced LIGO (aLIGO)~\cite{Abbott:2017xzg},
Virgo~\cite{virgo}, KAGRA~\cite{Akutsu:2018axf}, Einstein Telescope
(ET)~\cite{Sathyaprakash:2009xs,Moore:2014lga,ET_sense}, Cosmic
Explorer~\cite{Evans:2016mbw}), and space based interferometer
experiments
(LISA~\cite{Sathyaprakash:2009xs,Moore:2014lga,Audley:2017drz},
DECIGO~\cite{Seto:2001qf,Yagi:2011wg},
BBO~\cite{phinney2003big,Yagi:2011wg}).  Future measurements of
stochastic GWs can be a key to reveal the evolution history of the
Universe.

In this work, we focus on the relation between the induced GWs and an
early matter-dominated era (eMD era).  An eMD era is a period during
which the energy density of a massive field dominates the Universe
before the reheating.  Although the eMD effects on the induced GWs
have been discussed in
Refs.~\cite{Assadullahi:2009nf,Alabidi:2013wtp,Kohri:2018awv}, in our
accompanying paper~\cite{Inomata:2019zqy}, we find that if we
carefully take into account the evolution of the gravitational
potential, which is the source of the induced GWs, around the transition from an
eMD era to the radiation dominated era (RD era), the predictions for
the induced GWs can change.  In particular, we show that the induced GWs can
be significantly suppressed for a gradual transition, whose transition
timescale is comparable to the Hubble time at that time.  In some
cosmological scenarios (see Appendix~\ref{sec:model}) however, the
transition from an eMD era to the RD era is sudden, i.e. the timescale
of the transition is much shorter than the Hubble time at that time.
The purpose of this paper is to study the induced GWs in such sudden
transition cases.

The effects on the induced GWs of a sudden reheating mainly arise
during the RD era by the scalar perturbations that have already
entered the horizon during an eMD era.  Although GWs induced during an
eMD era have been studied in
Refs.~\cite{Assadullahi:2009nf,Alabidi:2013wtp,Kohri:2018awv}, GWs
induced during the RD era by the perturbations experiencing an eMD era
on subhorizon scales have not been investigated in the previous
studies.  However, we point out that, in sudden-reheating scenarios,
GWs induced during the RD era can be larger than GWs induced during an
eMD era by several orders of magnitude. Making use of this
enhancement, we might be able to determine the reheating temperature
by future GW detectors, as we discuss later.

We begin by reviewing the equations to calculate the induced GWs in
Sec.~\ref{sec:formalism}, and based on these equations we obtain the
power spectra for the induced GWs in Sec.~\ref{sec:gw_calculation}
assuming a sudden transition from an eMD era to the RD era. We explore
possibilities to determine the reheating temperature in
Sec.~\ref{sec:reheating _constraints} by making use of the enhancement
of the induced GWs associated with a sudden reheating, and
Sec.~\ref{sec:conclusions} is dedicated to conclusions. In Appendix
\ref{sec:model}, we discuss two models that can realize a sudden
transition from an eMD era to the RD era, and in Appendix
\ref{sec:analytic} we present approximate formulas for the induced GWs
in sudden-reheating scenarios.

%%%%%%%%%%%%%%%%%%%%%%%%%%%%%%
\section{Formulas for induced gravitational waves}
\label{sec:formalism}
%%%%%%%%%%%%%%%%%%%%%%%%%%%%%%

In the following, we briefly review the equations to calculate induced
GWs (see Ref.~\cite{Kohri:2018awv} for more details).  We assume that
an eMD era ended abruptly with the Universe entering into the RD era
at a conformal time $\eta = \eta_\text{R}$. Then the scale factor and
the conformal Hubble parameter are given by
\begin{align}
\frac{a(\eta)}{a(\eta_{\text{R}})} =& \begin{cases}
\left( \frac{\eta}{\eta_{\text{R}}} \right)^2          \\     
2 \frac{\eta}{\eta_\text{R}} - 1 
\end{cases},
&
\mathcal{H}(\eta)=& \begin{cases}
\frac{2}{\eta}                           &  \quad (\eta \leq \eta_\text{R}) \\
\frac{1}{\eta - \eta_{\text{R}}/2} & \quad  (\eta > \eta_\text{R})
\end{cases}.
\end{align}
We also assume that the curvature perturbations follow a Gaussian
distribution\footnote{GWs induced by the curvature perturbations with
  non-Gaussianity are discussed in
  Refs.~\cite{Nakama:2016gzw,Garcia-Bellido:2017aan,Cai:2018dig,Unal:2018yaa}}
and adopt the conformal Newtonian gauge\footnote{The gauge dependence
  of induced GWs is discussed in
  Refs.~\cite{Arroja:2009sh,Hwang:2017oxa}.} for simplicity.  Since we
focus on the effects of an eMD era, relevant to very small-scale
fluctuations
($k \gg k_\text{eq} = 0.0103\,
\text{Mpc}^{-1}$~\cite{Aghanim:2018eyx}),
in this paper we do not consider the enhancement of the induced GWs during
the late MD era ($z\lesssim 3400$)~\cite{Mollerach:2003nq,
  Baumann:2007zm}.

The energy density parameter of GWs per logarithmic interval in $k$ is given by
\begin{align}
\Omega_{\rm{GW}} (\eta, k) &= \frac{\rho_{\rm{GW}} (\eta, k) } { \rho_{\rm{tot}}(\eta)} \nonumber\\ 
&=
\frac{1}{24} \left( \frac{k}{a(\eta) H(\eta) } \right)^2 
\overline{\mathcal P_{h} (\eta,k)} ,
\label{eq:gw_formula}
\end{align}
where $\overline{\mathcal P_{h} (\eta,k)}$ is the time averaged power spectrum of GWs. 
It can be evaluated from the power spectrum ${\cal P}_\zeta$ of the
curvature perturbations by~\cite{Kohri:2018awv}
\begin{align}
\overline{\mathcal P_{h} (\eta,k)} = 4\int^\infty_0 \dd v &\int^{1+v}_{|1-v|} \dd u \left( \frac{4v^2 - (1+v^2 - u^2)^2}{4vu}  \right)^2 \nonumber \\
& \times \overline{I^2(u,v,k,\eta,\eta_\R) }\mathcal P_\zeta(u k) \mathcal P_\zeta(v k).
\label{eq:p_h_formula}
\end{align}
Here, $I(u,v,k,\eta,\eta_\R)$, describing the time dependence of GWs, is given by
\begin{align}
I(u,v,k,\eta,\eta_\R) =& \int^{x}_0 \dd \bar{x} \frac{a(\bar \eta)}{a(\eta)} k G_k(\eta, \bar \eta) f(u,v,\bar x, x_\R),
\label{eq:i_formula}
\end{align}
where $x$ and $x_\text{R}$ are defined as $x \equiv k\eta$ and $x_\text{R} \equiv k\eta_\text{R}$.
In this equation, $G_k$ is the Green's function being the solution of
\begin{align}
G_k''(\eta, \bar \eta) + \left( k^2 - \frac{a''(\eta)}{a(\eta)} \right) G_k(\eta, \bar \eta) = \delta (\eta - \bar \eta),
\label{eq:g_formula}
\end{align}
where a prime denotes differentiation with respect to $\eta$, not $\bar \eta$.
Note that the concrete expression of $G_k$ depends on the background evolution of the Universe, which is an eMD era or the RD era in our problem.
In addition, $f(u,v,\bar x, x_\R)$ is the source function defined as
\begin{widetext}
\begin{align}
f(u,v,\bar{x}, x_\R)=& \frac{3 \left( 2(5+3w) \Phi(u\bar{x})\Phi(v\bar{x})+4 \mathcal H^{-1} (\Phi'(u\bar{x})\Phi(v\bar{x})
 + \Phi(u\bar{x})\Phi'(v\bar{x})) 
+ 4 \mathcal H^{-2} \Phi'(u\bar{x})\Phi'(v\bar{x}) \right) }{25(1+w)},
\label{eq:f_def}
\end{align}
\end{widetext}
where $\omega = P/\rho$ is the equation-of-state parameter with $P$
and $\rho$ being the pressure and the energy density, respectively.
$\Phi$ is the transfer function of the gravitational potential, which
satisfies $\Phi(x \rightarrow 0, x_\R) \rightarrow 1$, and a prime
denotes differentiation with respect to the conformal time, that is,
$\Phi'(u \bar{x}) \equiv \partial \Phi(u \bar x)/\partial \bar \eta =
u k \,\partial \Phi(u\bar x)/\partial (u \bar x)$.
The second argument of $\Phi$ is abbreviated  in
Eq.~\eqref{eq:f_def} for compact notation.  $\Phi(u \bar{x})$
actually means $\Phi (u\bar{x}, u\bar{x}_\text{R})$, and 
$\Phi(v\bar{x})$ should be understood similarly.

The evolution equation for $\Phi$ is~\cite{Mukhanov:991646} 
\begin{align}
\Phi'' +  3(1+w)\mathcal H \Phi' + w k^2 \Phi = 0.
\label{eq:phi_evo_eq}
\end{align} 
By solving this equation, we find
\begin{align}
\Phi(x,x_\R) = \begin{cases}
1 & ( \text{for}\  x \leq x_\text{R}), \\
 A(x_\text{R}) \mathcal J(x) + B(x_\R) \mathcal Y(x)  & ( \text{for}\  x \geq x_\text{R}),
\end{cases}
\label{eq:phi_formula}
\end{align}
where we have dropped the decaying mode for $\eta < \eta_\R$. 
In this expression, $\mathcal J(x)$ and $\mathcal Y(x)$ are defined from the first and second spherical Bessel functions, $j_1(x)$ and $y_1(x)$, as
\begin{align}
\mathcal J(x) =& \frac{ 3\sqrt{3} \,  j_1\left(\frac{x-x_{\text{R}}/2}{\sqrt{3}} \right)}{x-x_{\text{R}}/2}, \\ 
\mathcal Y(x) =& \frac{3\sqrt{3} \, y_1\left( \frac{x - x_{\text{R}}/2}{\sqrt{3}} \right)}{x- x_{\text{R}}/2}, 
\label{eq:c_d_formula}
\end{align}
and the coefficients $A(x_\R)$ and $B(x_\R)$ are determined so that $\Phi(x)$ and $\Phi'(x)$ are continuous at $x=x_\R$:
\begin{align}
\label{eq:a_formula}
A(x_\R) &= \frac{1}{\mathcal J(x_\R) - \frac{\mathcal Y(x_\R)}{\mathcal Y'(x_\R)} \mathcal J'(x_\R)}, \\
\label{eq:b_formula}
B(x_\R) &= - \frac{\mathcal J'(x_\R)}{\mathcal Y'(x_\R)} A(x_\R).
\end{align}
In Appendix~\ref{app:sudden_reh_model}, we introduce a model realizing a sudden-reheating transition and check that the above analytic expression for $\Phi$ with these connection conditions coincides well with the numerical solution for $\Phi$ calculated for that model. 

We can reexpress Eq.~(\ref{eq:i_formula}) as in Ref.~\cite{Kohri:2018awv} with a slight refinement of the time dependence of the scale factor:
\begin{align}
I(u,v,x,x_\R) =& \int^{x_\text{R}}_0 \dd \bar{x} \left( \frac{1}{2(x/x_\R) -1} \right) \left( \frac{\bar x}{x_\text{R}} \right)^2 \nonumber \\
&\qquad \qquad
\times k G_k^{\text{eMD}\rightarrow \text{RD}}(\eta, \bar \eta) f(u,v,\bar x,x_\R) \nonumber\\
&
+ \int^x_{x_\text{R}} \dd \bar x \left( \frac{2(\bar x/x_\R) -1}{2(x/x_\R) -1} \right) \nonumber \\
& \qquad \qquad 
\times 
k G^{\text{RD}}_k (\eta, \bar \eta) f(u,v,\bar x,x_\R) \nonumber  \\
\equiv& I_\text{eMD}(u,v,x,x_\R) + I_\text{RD}(u,v,x,x_\R),
\label{eq:i_formula_emd}
\end{align}
where $I_\text{eMD}$ and $I_\text{RD}$ represent the contributions
from GWs induced during an eMD era and the RD era, respectively. See our accompanying paper~\cite{Inomata:2019zqy} for the concrete expressions of the Green functions. We approximate $\overline{I^2(u,v,x,x_\R)}$ in Eq.~(\ref{eq:p_h_formula}) as
\begin{align}
\overline {I^2(u,v,x,x_\R)} \simeq \overline{I_\text{eMD}^2(u,v,x,x_\R)} + \overline{I_\text{RD}^2(u,v,x,x_\R)}.
\end{align}
Correspondingly, we approximately split $\Omega_\text{GW}$ into two parts as $\Omega_\text{GW} \simeq \Omega_\text{GW,RD} + \Omega_\text{GW,eMD}$, 
where $\Omega_\text{GW,RD}$ and $ \Omega_\text{GW,eMD}$ are calculated from $\overline{I^2_\text{RD}}$ and  $\overline{I^2_\text{eMD}}$, respectively.

The
analytic formulas for $I_\text{eMD}$ and $I_\text{RD}$ are derived in
Ref.~\cite{Kohri:2018awv}. In this reference, we adopted an implicit
assumption that GWs induced during the RD era by the perturbations
having entered the horizon during an eMD era, which we focus on in
this work, are subdominant compared to the GWs 1)
which are induced during the eMD era and 2) which are induced by the
perturbations entering the horizon after the reheating.
However, this assumption is not true in realistic situations.  In our
accompanying paper~\cite{Inomata:2019zqy}, we consider a gradual
reheating transition and show that the contributions from
$I_\text{RD}$ play an important roll for the suppression of induced
GWs.  In addition, in a sudden-reheating scenario, the dominant contribution comes from $I_\text{RD}$ as we will show in
Sec.~\ref{sec:gw_calculation}.

During the RD era, the gravitational potential, the source of GWs,
decays on subhorizon scales, and therefore $\Omega_{\text{GW}}$
becomes constant after the gravitational potential has sufficiently
decayed.  Here, we define $\eta_c$ as the moment when
$\Omega_\text{GW}$ becomes constant.
Note that since we focus on small scales, where the effects of an eMD
era may leave traces, $\eta_c$ is well before the standard
matter-radiation equality time.  Taking into account the evolution of
GWs during the late MD era and the change in the effective
relativistic degrees of freedom, we can write the current energy
density parameter $\Omega_\text{GW}(\eta_0,k)$ as~\cite{Ando:2018qdb}
\begin{align}
%\Omega_{\rm{GW}} (\eta_0, k)  = 0.83 \left( \frac{g_c}{10.75} \right)^{-1/3} \Omega_{\text{r},0} \Omega_{\rm{GW}}(\eta_c, k),
\Omega_{\rm{GW}} (\eta_0, k)  = 0.39 \left( \frac{g_c}{106.75} \right)^{-1/3} \Omega_{\text{r},0} \Omega_{\rm{GW}}(\eta_c, k),
\label{eq:latetime}
\end{align}
where $\Omega_{\text{r},0}$ is the current value of the energy density
parameter for radiation. In this paper, we denote the effective
relativistic degrees of freedom by $g$, and $g_c$ in this equation is
its value at $\eta = \eta_c$.

%%%%%%%%%%%%%%%%%%%%%%%%%%%%%%
\section{Calculations of induced gravitational waves}
\label{sec:gw_calculation}
%%%%%%%%%%%%%%%%%%%%%%%%%%%%%%
Using the above equations and the analytic formulas in Ref.~\cite{Kohri:2018awv}, we calculate induced GWs.
To be specific, we assume the following power spectrum of the curvature perturbation:
\begin{align}
\mathcal P_\zeta (k) =  \Theta(k_\text{max} - k) A_s \left( \frac{k}{k_*} \right)^{n_s-1},
\label{eq:pzeta_def}
\end{align}
where $A_s$ is the amplitude at the pivot scale $k_*$, $n_s$ is the tilt of the power spectrum, and $\Theta$ is the Heaviside step function. 
We introduce $k_\text{max}$ as the cutoff scale of the power spectrum. Since matter density perturbations grow in proportion to the scale factor during a MD era, perturbations may enter into the non-linear regime, depending on the amplitude of primordial fluctuations and the duration of an eMD era. If perturbations remain in the linear regime during an eMD era, the cutoff scale is the Hubble radius at the beginning of an eMD era. On the other hand if such nonlinearities arise, $k_{\text{max}}$ should be chosen as the wavenumber of perturbations that are entering the non-linear regime at the end of an eMD era, since our formalism is based on the linear theory.\footnote{
	There are works discussing GWs induced by non-linear perturbations, though the results inevitably involve some uncertainties~\cite{Jedamzik:2010dq,Jedamzik:2010hq}.
}
In this case, the smallest scale on which we can apply the linear theory is approximated as~\cite{Inomata:2019zqy}
\begin{align}
k_\text{NL} \sim 470 /\eta_\text{R} .
\label{eq:nl_scale}
\end{align}
Hence, to ensure the validity of our analysis, we limit our analysis to cases with $k_\text{max} \leq 450 /\eta_\text{R}$. That is, we do not take into account GWs that are generated from non-linear perturbations, and this means our analysis would lead to conservative estimations of the GW spectrum. 

\begin{figure}
\centering
\includegraphics[width=0.48 \textwidth]{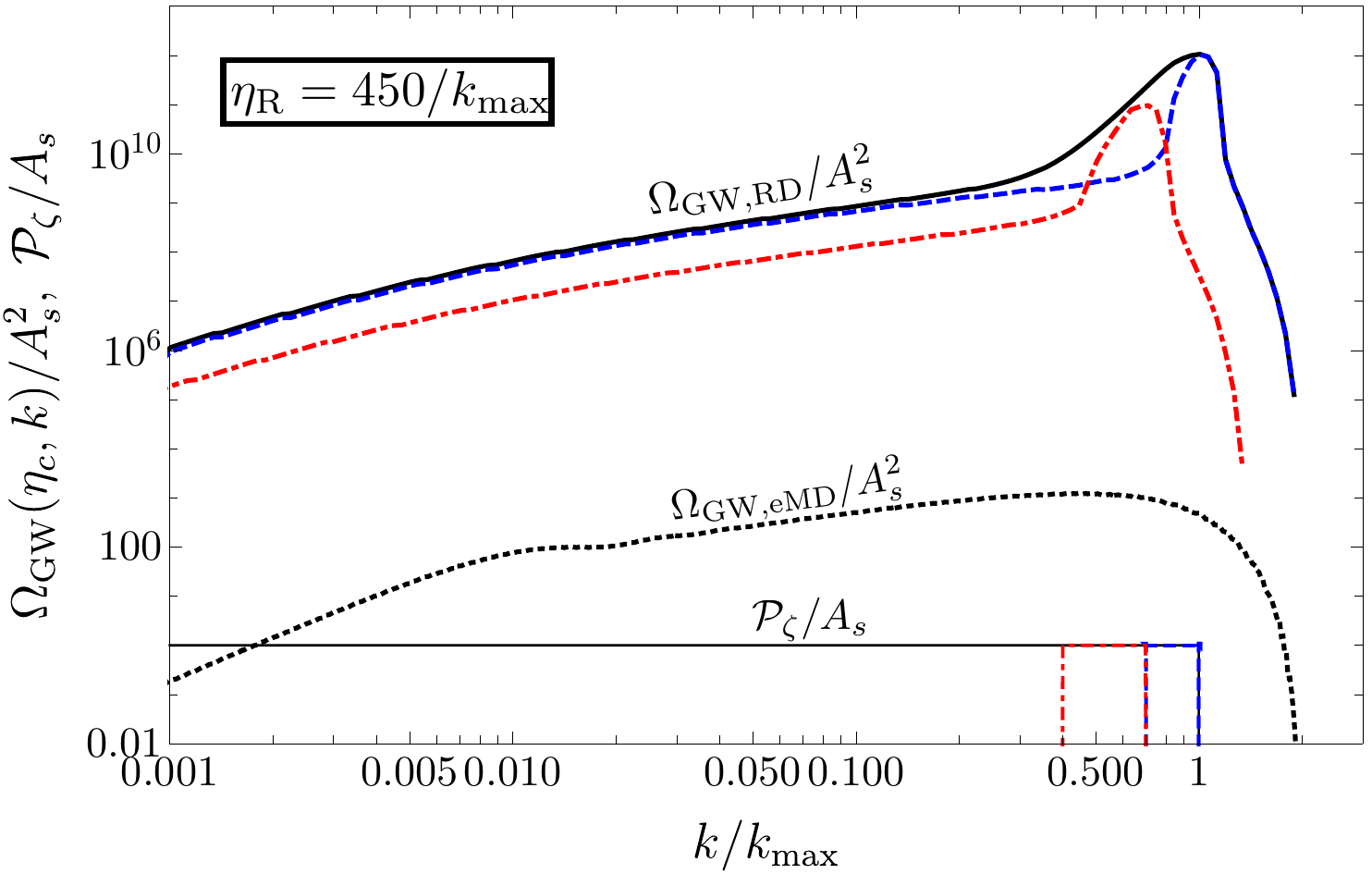}
\caption{Energy density parameters of GWs, for each logarithmic interval of wavenumber, induced during the RD era ($\Omega_\text{GW,RD}(\eta_c,k)$) and during an eMD era ($\Omega_\text{GW,eMD}(\eta_c,k)$),
as well as the power spectrum $\mathcal P_\zeta(k)$ of curvature perturbations. They are normalized by $A_s^2$ or $A_s$, respectively, and
we take $\eta_\text{R} = 450/ k_\tmax$.
The black lines are derived from $\mathcal P_\zeta(k) = \Theta(k_\text{max} - k) A_s$.
For comparison, the blue and red lines are also shown, which are derived from $\mathcal P_\zeta(k) = \Theta(k_\text{max} - k)\Theta(k - 0.7 k_\text{max}) A_s$ and $\mathcal P_\zeta(k) = \Theta(0.7 k_\text{max} - k)\Theta(k - 0.4 k_\text{max}) A_s$, respectively.
}
\label{fig:emd_gw_profiles}
\end{figure}

Figure~\ref{fig:emd_gw_profiles} shows the scale dependence of the spectrum of induced GWs.
In this figure, we take $n_s = 1$ for simplicity.  We can see that
$\Omega_\text{GW,RD}$ is much larger than $\Omega_\text{GW,eMD}$.
This is because the GWs induced during the RD era by the subhorizon
perturbations that entered the horizon during an eMD era, neglected
in the previous works but taken into account in this work, are
significant.  For comparison, we also plot the GW spectra induced by
the power spectra of
$\mathcal P(k) = \Theta(k_\text{max}-k) \Theta(k - 0.7
k_\text{max})A_s$
and
$\mathcal P(k) = \Theta(0.7 k_\text{max}-k) \Theta(k - 0.4
k_\text{max})A_s$ with blue and red lines.
As shown in the figure, the contributions from the smallest scales
(blue dashed line) are the dominant contributions to the total
spectrum (solid black line) except for the large-scale-side tail of
the sharp peak ($0.3 \lesssim k/k_\text{max} \lesssim 1$).  This sharp peak is due to the resonance effect, which
is a characteristic feature of GWs induced during the RD
era~\cite{Ananda:2006af}, when the gravitational potential oscillates.
The tail of the sharp peak is formed by the envelope of the resonance
effects on these scales (see the red dot-dashed line).  In this way,
the spectrum of the induced GWs is produced dominantly by the smallest
scales, and the resonant amplification plays a key role.  This
understanding becomes clearer in Appendix~\ref{sec:analytic}, where we
derive approximate analytic formulas for induced GWs for sudden-reheating scenarios.
On much larger scales, the contributions from the perturbations
entering the horizon after the reheating dominate induced GWs, whose
spectrum becomes scale invariant
$\Omega_\text{GW}(\eta_c,k) \simeq 0.8222
A_s^2$~\cite{Kohri:2018awv}.
This can be observed in the GW spectrum for $k<10^9$Mpc$^{-1}$ in
Fig.~\ref{fig:emd_induced_gws}, though in that figure a slightly
scale-dependent primordial spectrum is assumed, leading to a slight
scale dependence of $\Omega_{\text{GW}}$.

The main reason why induced GWs are enhanced is that the gravitational
potential $\Phi$ with large $k$ ($\gg 1/\eta_\R$) is constant until
$\eta = \eta_\R$ and, after the reheating, it oscillates with the
timescale $\sim 1/k$, much shorter than its decay timescale
$\sim \eta_\R$. 
Due to the fast oscillations of perturbations with
unsuppressed amplitudes, which remained constant until the moment of the reheating, induced
GWs are significantly enhanced.\footnote{Although the perturbations entering the horizon during the RD era also oscillate with the timescale much shorter than their decay timescale well after (not soon after) the horizon entry, the amplitudes of the perturbations start to decay soon after the horizon entry, unlike during an eMD era, and therefore the enhancement is not caused by the perturbations entering the horizon during the RD era. }
Note that the dominant contributions come from the last term in Eq.~(\ref{eq:f_def}), which involves two time derivatives of the gravitational potential.
This is because, at the beginning of the RD era, the last term can be approximated as $\mathcal H^{-2} \Phi'\Phi' \sim (k\eta_\R)^2\Phi^2 \gg \Phi^2$ for the perturbations that entered the horizon well before the sudden reheating. In other words, the factor $(k\eta_\R)^2$ in the source term and the amplitude of $\Phi$ that remained constant until the reheating are the main causes for the enhancement.

In addition to numerical solutions, we also obtain approximate
analytic formulas of $\Omega_{\text{GW,RD}}$ in
Appendix~\ref{sec:analytic}, with $\Omega_{\text{GW,RD}}$ given by the
sum of Eqs.~\eqref{eq:Omega_GW_large-scale} and
\eqref{eq:Omega_GW_resonance}.  Using these expressions, the GW
spectrum is roughly expressed as
\begin{align}
\frac{\Omega_{\text{GW}}(\eta_c,k)}{A_{\text{s}}^2} \simeq & 
\begin{cases}
0.8 & (x_{\text{R}} \lesssim 150 x_{\text{max,R}}^{-5/3}) \\
3 \times 10^{-7} x_{\text{R}}^3 x_{\text{max,R}}^5  &   (150 x_{\text{max,R}}^{-5/3} \lesssim x_{\text{R}} \ll 1) \\
1 \times 10^{-6} x_{\text{R}} x_{\text{max,R}}^5 & (1 \ll x_{\text{R}} \lesssim x_{\text{max,R}}^{5/6}) \\
7 \times 10^{-7}   x_{\text{R}} ^7 &     (x_{\text{max,R}}^{5/6} \lesssim x_{\text{R}}   \lesssim x_{\text{max,R}})  \\
\text{(sharp drop)} & (x_{\text{max,R}} \lesssim x_{\text{R}} \leq 2 x_{\text{max,R}})
 \end{cases},  \label{eq:Omega_GW_rough_behavior}
\end{align}
neglecting a logarithmic factor for the second line.

%%%%%%%%%%%%%%%%%%
%%%%%%%%%%%%%%%%%%

%%%%%%%%%%%%%%%%%%%%%%%%%%%%%%
\section{Determination of reheating temperature}
\label{sec:reheating _constraints}
%%%%%%%%%%%%%%%%%%%%%%%%%%%%%%

\begin{figure}
\centering
\includegraphics[width=0.48 \textwidth]{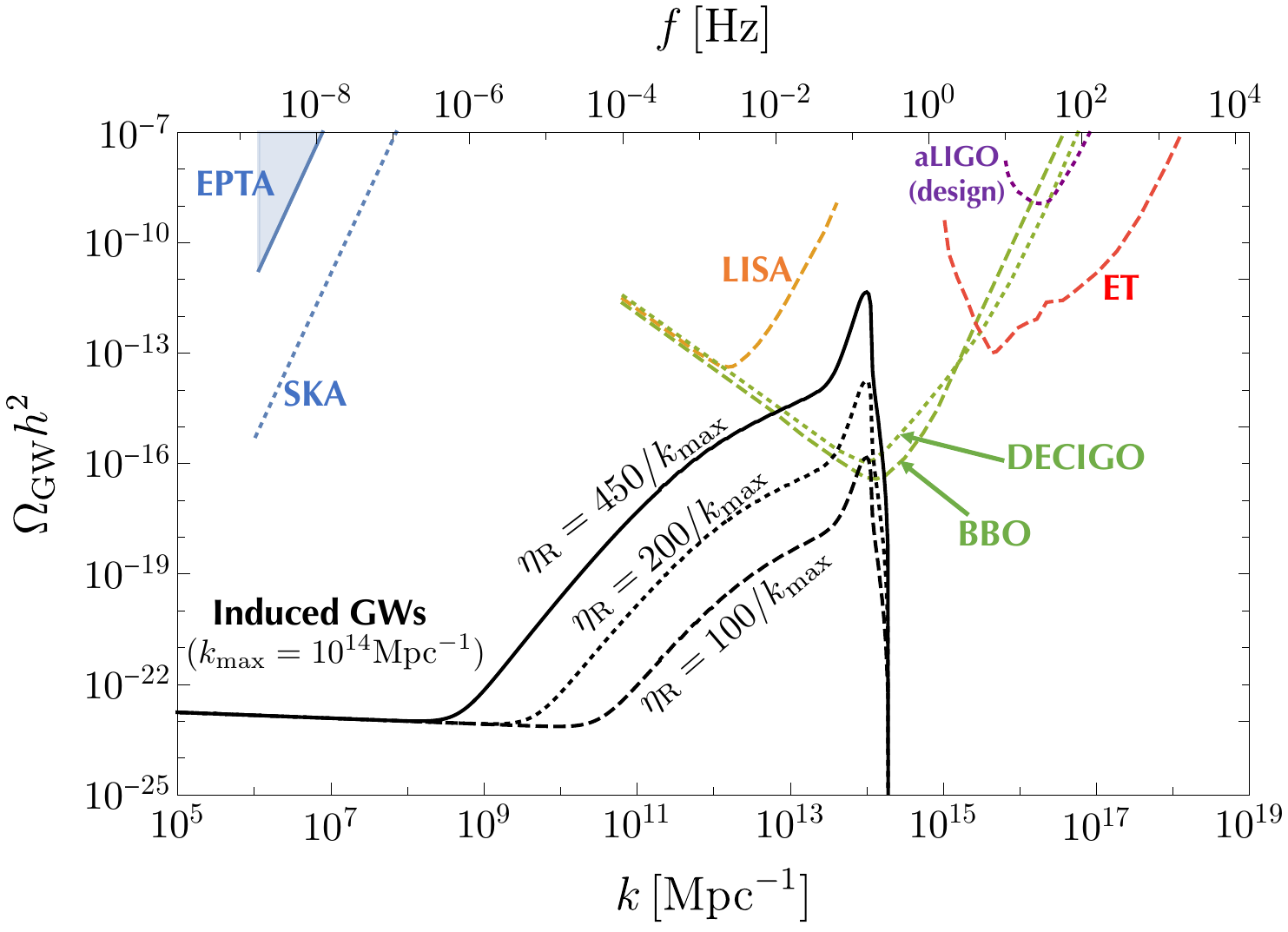}
\caption{Effective sensitivities to stochastic GWs of current and future experiments.	
	Note that we plot $\Omega_\text{GW,eff} h^2 /\sqrt{T_\text{obs} f/10}$ as a sensitivity curve for each experiment.
	We consider the same experiments and take the same parameters as in Ref.~\cite{Inomata:2018epa}.
	In particular, we take the same parameter sets of observation time for each experiment: $T_\text{obs} = 18$ years for EPTA, $T_\text{obs} = 20$ years for SKA, and $T_\text{obs} = 1$ year for the other experiments.
	The shaded regions have already been excluded by the existing observational data.
	See Ref.~\cite{Inomata:2018epa} for more details about the sensitivity curve of each experiment.
	Black lines show the energy density parameters of the GWs induced by the power spectrum of $\mathcal P_\zeta(k) = 2.1\times 10^{-9} (k/0.05\,\text{Mpc}^{-1})^{-0.04} \Theta(k_\tmax - k)$.
	We take $k_\tmax = 10^{14}$Mpc$^{-1}$ for all these three lines and 
	 $\eta_\R = 450 / k_\tmax$, $\eta_\R = 200 / k_\tmax$ and $\eta_\R = 100 / k_\tmax$ for each line, respectively.
}
\label{fig:emd_induced_gws}
\end{figure}

In the previous section, we have shown that the induced GWs can be
much larger than those previously
reported~\cite{Kohri:2018awv,Alabidi:2013wtp,Assadullahi:2009nf}.  In
the following, we consider the GWs induced by the almost
scale-invariant power spectrum, given in Eq.~(\ref{eq:pzeta_def}),
with $A_s=2.1\times 10^{-9}$, $k_* = 0.05$\,Mpc$^{-1}$, and
$n_s = 0.96$~\cite{Aghanim:2018eyx}.  Figure~\ref{fig:emd_induced_gws}
shows the sensitivity curves of current and future GW experiments and
plots for $\Omega_{\text{GW}}$ of the GWs induced by this power spectrum
with $k_\tmax = 10^{14}$Mpc$^{-1}$.  This figure shows that the induced
GWs associated with a sudden transition from an eMD era to the RD era
could in principle be observable by future projects.  Since the height
and scale of the peak are determined by the scale of the reheating and
the cutoff $k_\tmax$, we discuss what range of the reheating
temperature could be probed by future observations searching for GWs.

%%%%%%%%%%%%%%%%%%
%%%%%%%%%%%%%%%%%%

We adopt an analysis similar to that in our previous
paper~\cite{Inomata:2018epa} (see also Ref.~\cite{Thrane:2013oya}).
We use the signal-to-noise ratio $\rho$ for GW interferometers given
by~\cite{Thrane:2013oya}.
\begin{align}
\rho = \sqrt{2T_\text{obs}} \left[ \int^{f_\text{max}}_{f_\text{min}} \dd f \, \left(\frac{\Omega_\text{GW}(f)}{\Omega_\text{GW,eff}(f)}\right)^2 \right]^{1/2}.
\label{eq:rho_def}
\end{align}
Here, $T_\text{obs}$ is the observation time, and $(f_\text{min}, f_\text{max})$ is the range of observable 
frequencies for each project. 
$\Omega_\text{GW,eff}$ is the
effective sensitivity curve, which is calculated for each project (see
Ref.~\cite{Inomata:2018epa} for detail).  For pulsar timing array
(PTA) observations, we use
~\cite{Anholm:2008wy,Siemens:2013zla,Chamberlin:2014ria}
\begin{align}
\rho = &\sqrt{2T_\text{obs}} \left(\sum_{I=1}^M  \sum_{J>I}^M \chi^2_{IJ} \right)^{1/2} \nonumber \\
&\times \left[ \int^{f_\text{max}}_{f_\text{min}} \dd f \, \left(\frac{\Omega_\text{GW}(f)}{\Omega_\text{n} (f) + \Omega_\text{GW}(f)}\right)^2 \right]^{1/2}.
\label{eq:pta_snr}
\end{align}
In this expression, $M$ is the number of the observed pulsars, $\chi_{IJ}$ is the Hellings and Downs coefficient, and $\Omega_\text{n}$ is the energy density parameter for noise of each pulsar. We take the same parameters and noise power spectrum as in Ref.~\cite{Inomata:2018epa} assuming that the noise is dominated by the white timing noise~\cite{Thrane:2013oya}.

\begin{figure}
\centering
\includegraphics[width=0.48 \textwidth]{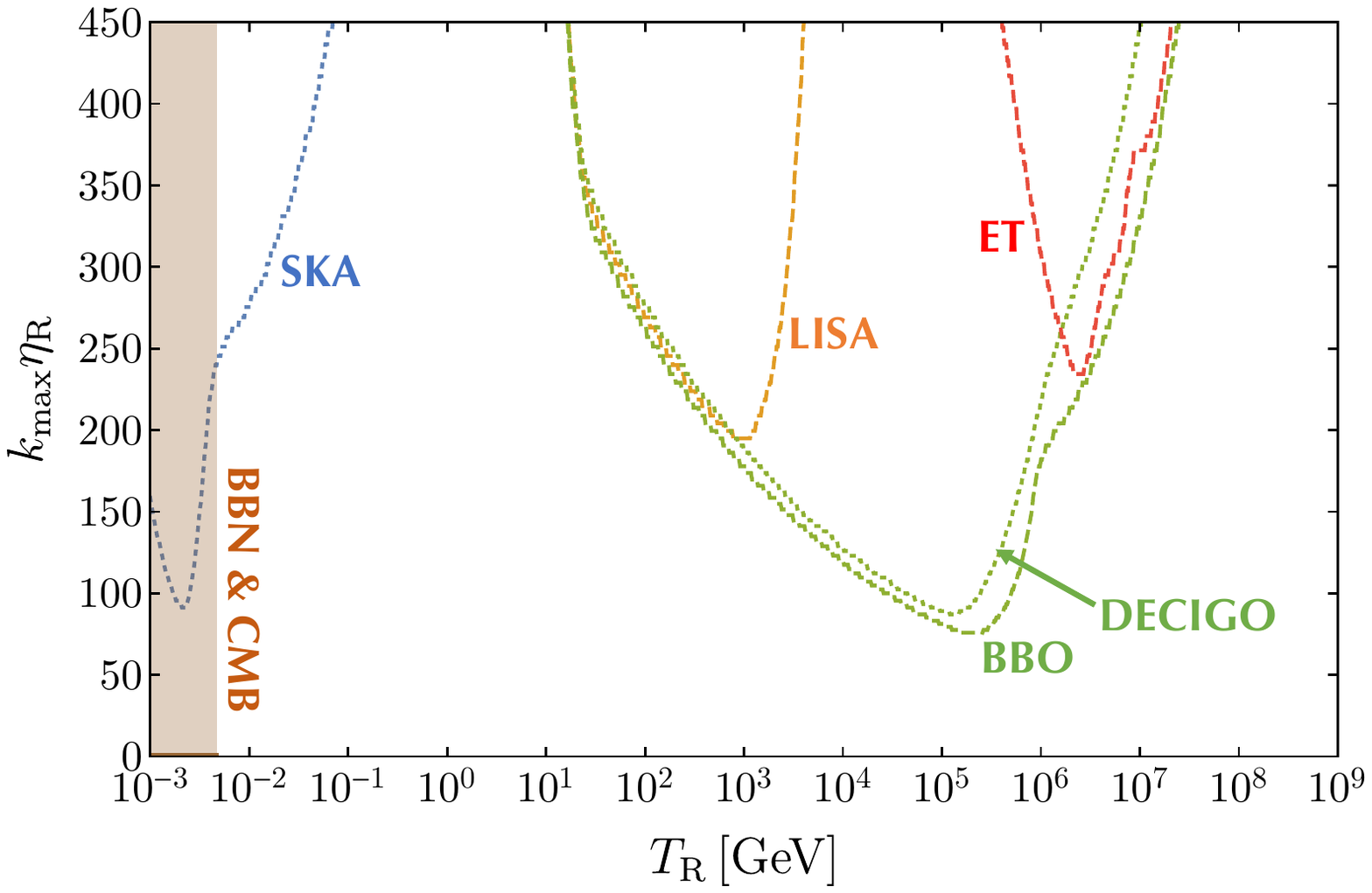}
\caption{\small	
	Relation between the cutoff scale multiplied by $\eta_\R$ and the reheating temperature that can be probed by the future observations.
	We take $T_\text{obs} = 20$ yr for SKA and $T_\text{obs}=1$ yr for the others.
	The curves correspond to the values of $k_\text{max}\eta_\R$ required to reach the signal-to-noise ratio of unity ($\rho = 1$) for the experiments at each reheating temperature.
	The brown shaded region is already excluded by the big bang nucleosynthesis and the Planck data~\cite{deSalas:2015glj}.
}
\label{fig:reh_const_kmax}
\end{figure}

Using the effective sensitivity curves in Fig.~\ref{fig:emd_induced_gws}, we derive the cutoff scale to give $\rho=1$ for each project and reheating temperature.
The numerical results are shown in Fig.~\ref{fig:reh_const_kmax}. 
When we derive the curves, we use the approximation formulas given in Eqs.~(\ref{eq:Omega_GW_large-scale}), (\ref{eq:Omega_GW_resonance}), and (\ref{eq:omega_gw_approximation}) to save the computational time.\footnote{
For simplicity, we use the formulas for the scale-invariant spectrum, given in Eqs.~(\ref{eq:Omega_GW_large-scale}), (\ref{eq:Omega_GW_resonance}), and (\ref{eq:omega_gw_approximation}), instead of those for the power-law spectrum, given in Eqs.~(\ref{eq:omega_tilt_ls})-(\ref{eq:omega_tilt_res}).
Since the enhancement of the induced GWs is mainly caused by the perturbations on the smallest scales ($k\sim k_\text{max}$) and the tilt of the power spectrum is small, we can approximately use the formulas for the scale-invariant spectrum whose amplitude is given by $A_s (k_\text{max}/k_*)^{n_s-1}$.
We have also numerically checked that the effect of the tilt with $n_s = 0.96$ around the smallest scale on the enhanced GW spectrum is negligible.}
In this figure, we take $T_\text{obs} = 20$ years for SKA and $T_\text{obs}=1$ year for the other projects and assume no foreground for simplicity.
When we obtain these plots, we have used the following relation between the conformal time and the temperature~\cite{Inomata:2018epa}\footnote{
To derive Eq.~(\ref{eq:eta_t_rel}), we use the relation $a_\text{eq}H_\text{eq} = k_\text{eq} (= 0.0103\, \text{Mpc}^{-1}$)~\cite{Aghanim:2018eyx,Ade:2015xua}.
This relation corrects the factor given in Ref.~\cite{Inomata:2018epa}.
}
\begin{align}
	\label{eq:eta_t_rel}
   & \frac{aH}{a_\text{eq} H_\text{eq}} = \frac{1}{\sqrt{2}}  \left(\frac{g_{s,\text{eq}}}{g_s}\right)^{1/3} \left( \frac{g}{g_\text{eq}} \right)^{1/2} \frac{T}{T_\text{eq}}, \nonumber \\   \Rightarrow \ 
   & \frac{\eta_\R}{10^{-14} \text{Mpc}} = \left( \frac{g_s}{106.75} \right)^{1/3} \left( \frac{g}{106.75} \right)^{-1/2} \nonumber \\
   & \qquad\qquad\qquad\qquad\qquad
   \times  \left(\frac{T_\R}{1.2 \times 10^{7}\text{GeV}}\right)^{-1},
\end{align}
where the subscript ``eq'' means the value at the late
matter-radiation equality ($z\sim3400$), and $g_s$ is the effective
degrees of freedom for the entropy density.  Note again that the peak
scale of the induced GWs corresponds to $k\sim k_\tmax$, not
$k\sim 1/\eta_\R$.
Figure ~\ref{fig:reh_const_kmax} shows that, in the case of $k_\text{max}\eta_\R = 450$, the ranges of reheating temperature that future observations could investigate are $T_\R \lesssim 7\times 10^{-2}\, \GeV$ for SKA, $20\,\GeV \lesssim T_\R \lesssim 4\times 10^3\,\GeV$ for LISA, $20\,\GeV \lesssim T_\R \lesssim 1 \times 10^7\,\GeV$ for DECIGO, $20\,\GeV \lesssim T_\R \lesssim 2 \times 10^7\,\GeV$ for BBO, and $4\times 10^5\,\GeV \lesssim T_\R \lesssim 2\times 10^7\,\GeV$ for ET.

%%%%%%%%%%%%%%%%%%
%%%%%%%%%%%%%%%%%%

%%%%%%%%%%%%%%%%%%%%%%%%%%%%%%
\section{Conclusions}
\label{sec:conclusions}
%%%%%%%%%%%%%%%%%%%%%%%%%%%%%%

We have studied the effects of an eMD era on the induced GWs.  In
particular, we have focused on a sudden-reheating scenario, in which
the reheating completes on a timescale much shorter than the Hubble
time at that time.  Then, we have found that the induced GWs can be
significantly enhanced in such a scenario. The main contributions to
the enhanced GWs come from the GWs induced during the RD era by the
perturbations that entered the horizon during the eMD era.  This is
due to the fast oscillations of the perturbations after the sudden
transition.  This enhancement is qualitatively opposite to the
suppression of induced GWs in gradual-reheating scenarios, which we
report in our accompanying paper~\cite{Inomata:2019zqy}.  This means
that the eMD effects on the induced GWs strongly depend on how the
reheating takes place.

We have also numerically calculated the induced GWs with realistic
power spectra of curvature perturbations and discussed possibilities
of determining the reheating temperature observationally for sudden-reheating scenarios.
We have found that if an eMD era lasts for
$224 \eta_\text{eMD,start}$, where $\eta_\text{eMD,start}$ is the
conformal time at the start of the eMD era,\footnote{Note that the
  wavenumber corresponding to the horizon scale at
  $\eta_\text{eMD,start}$ satisfies $k_\text{eMD,start} \eta_\R= 450$
  because
  $k_\text{eMD,start} = a_\text{eMD,start} H_\text{eMD,start} =
  2/\eta_\text{eMD,start}$
  and $\eta_\R = 225 \eta_\text{eMD,start}$ in this case (the duration is $\eta_\text{R} - \eta_\text{eMD,start} = 224 \eta_\text{eMD,start}$).} the
reheating temperatures in the range
$T_\R \lesssim 7 \times 10^{-2}$GeV or
$20 \, \text{GeV} \lesssim T_\R \lesssim 2\times 10^7 \, \text{GeV}$
could be probed by future GW projects, such as SKA, LISA, DECIGO, BBO,
and ET.  Note that, if an eMD era starts right after inflation era,
the duration of the eMD era of $\mathcal O(100) \eta_\text{eMD,start}$
corresponds to $\rho^{1/4}_\text{inf}/T_\R \sim \mathcal O(10^3)$,
where $\rho_\text{inf}$ is the energy density during the inflation
era. 

Since the enhancement of the induced GWs we find is so significant,
one may wonder if it is consistent with some physical requirements.
Hence, we briefly mention some consistency checks. 
Let us first discuss energy conservation and backreaction.
Note that the dominant part of the energy density of the induced GWs is generated soon after the reheating transition by the short-wavelength modes at around $k \sim k_\text{max}$, and the enhancement is stronger for $k_\text{max}$ closer to the non-linear scale $k_\text{NL}$.  By definition, the energy density of density perturbations at scales around $k_\text{NL}$ is comparable to the energy density of the homogeneous component. On the other hand, in Fig.~\ref{fig:emd_gw_profiles}, we have seen that the energy density of the induced GWs is $\Omega_\text{GW}(\eta_c) \sim 10^{12} A_s^2 \sim \mathcal O(10^{-6})$ even for $k_\text{max} \eta_\text{R} = 450$ ($k_\text{max} \sim k_\text{NL}$). The smallness of $\Omega_\text{GW}$ implies that the energy density of the induced GWs is much smaller than both the energy density of its source, namely the density perturbations, and the homogeneous component.  Thus, a
backreaction of the GW production to the thermal history of the
Universe would be negligible.

Another concern may be whether or not GWs induced at third order in
scalar perturbations are negligible, given the fact that the
second-order contributions have turned out to be significant.
In other words, one might wonder whether or not sources coming from third order scalar perturbations appear with many derivatives, %e.g. $\mathcal H^{-5} \Phi''' \Phi' \Phi'$, 
which can be larger than the dominant source in second order scalar perturbations, $\mathcal H^{-2} \Phi'\Phi'$, even when the perturbations are linear $(k\eta_\R)^2 \Phi \lesssim 1$.
Complete evaluations of the third-order contributions are much more
complicated than those for the second-order analysis. Thus, we
estimate their orders of magnitudes by listing up possible third-order
terms that appear in the evolution equation for tensor perturbations
and are consistent with general covariance and the transverse
traceless condition.  Most of them turned out to be negligible
provided that we are in the linear regime, where the density
perturbations are less than unity. However, we need to carefully
evaluate the contributions involving the first-order scalar
perturbations multiplied by the second-order vector perturbations,
which are sourced by first-order scalar perturbations at second order,
similarly to the induced GWs we have studied. That is because 
 the power spectrum of second-order vector perturbations is
larger than that of second-order tensor
perturbations~\cite{Saga:2015apa}.  Full evaluations of these
contributions are beyond the scope of this work, but we expect that
the third-order GWs can be subdominant as long as
$k_{\text{max}} \ll k_{\text{NL}}$.

In this paper, we do not take into account the gauge dependence of
the induced GWs~\cite{Hwang:2017oxa}, GW foregrounds due to other
astrophysical as well as cosmological sources, and the possible
contributions from non-linear perturbations~\cite{Jedamzik:2010dq,
  Jedamzik:2010hq}, which can arise depending on the length of the
eMD era and the amplitude of the small-scale primordial
fluctuations. Although these issues remain to be investigated, our
result shows that observations of GWs could possibly reveal the
reheating history of the Universe in the near future.

\acknowledgments 

KI and TN thank KEK, Johns Hopkins University and Research Center for
the Early Universe, University of Tokyo, for hospitality received
during this work. 
KI acknowledges Tomohiro Fujita, Teruaki Suyama, and Masahide Yamaguchi for useful comments.
We thank Kai Murai for pointing out the necessity of the fudge factor $Y$ in Eq.~(B7).  (This factor was accidentally already included in our plots in the previous versions.)
We are also grateful to Hanwen Tai for pointing out an error in our numerical calculation.
This work was supported in part by World Premier
International Research Center Initiative (WPI Initiative), MEXT,
Japan, the JSPS Research Fellowship for Young Scientists (KI and TT),
JSPS KAKENHI Grants No.~JP18J12728 (KI), No.~JP17H01131 (KK), and No.~JP17J00731 (TT), MEXT
KAKENHI Grants No.~JP15H05889 (KK), No.~JP18H04594 (KK) and
No.~JP19H05114 (KK), and Advanced Leading Graduate Course for Photon
Science (KI).

\appendix

\section{A model that realizes a sudden-reheating transition} \label{sec:model}
\label{app:sudden_reh_model}

In this Appendix, we build a concrete model in which the reheating
happens in a timescale much shorter than the Hubble time at that time.
This ensures that $\Phi$ does not decay during the reheating
transition, leading to the enhancement of induced GWs, reported in
this paper, in contrast to the suppression of induced GWs for a
gradual transition, reported in our accompanying paper~\cite{Inomata:2019zqy}.

Our idea for a sudden reheating is to initially block the decay of the
field $\phi$, dominating the energy density in the eMD era, into
relativistic daughter particles, collectively denoted by $\chi$, for some reason
related to kinematics or symmetry, and then to remove the cause of the
blocking in a dynamical manner.  
For this purpose, we introduce a
field $\tau$, which dynamically triggers the decay of $\phi$ into
$\chi$. We dub such a field $\tau$ `triggeron'.  In the models we
discuss below, the mass of $\chi$ is dependent on the field value of
$\tau$ and a quick change of that field value causes a sudden decay of
$\phi$ to $\chi$, which we identify as a sudden reheating.

\subsection{A scenario for a sudden reheating triggered by a fast rolling field }

The main ideas of this model are as follows.  At first, the initial
triggeron value is sufficiently large so that the decay of $\phi$ into
two $\chi$ particles is kinematically forbidden.  When the Hubble
parameter becomes comparable to the triggeron mass $m$, the triggeron
starts to roll down its potential quickly, and it passes through some
critical value at which the decay channel of $\phi$ opens.  If the
decay rate is much larger than the Hubble scale, the reheating
transition completes quickly.

We consider a simple model that involves three canonically normalized real scalar fields $\phi$, $\tau$, and $\chi$ to demonstrate the ideas. One can easily generalize this model by considering e.g. complex scalar fields, fermions, or gauge bosons.
The Lagrangian density we assume is
\begin{align}
\mathcal{L}=& -\frac{1}{2}  \partial^\mu \phi \partial_\mu \phi  -\frac{1}{2}  \partial^\mu \chi \partial_\mu \chi  -\frac{1}{2}  \partial^\mu \tau \partial_\mu \tau  -V, \\
V=& \frac{1}{2} M^2 \phi^2 + \frac{1}{2}m^2 \tau^2 + \frac{\lambda}{4} \tau^2 \chi^2 + \frac{c}{2}M \phi \chi^2,
\end{align}
where $M$ and $m$ denote the masses of $\phi$ and $\tau$, respectively, satisfying $M^2 \gg m^2$, and $\lambda$ and $c$ are dimensionless coupling constants.  The third term in the potential can be interpreted as the $\tau$-dependent mass term for $\chi$, and the last term provides the decay channel of $\phi$ into 2 $\chi$ particles, once the decay becomes kinematically allowed.
The decay rate of $\phi$ into 2 $\chi$ particles is 
\begin{align}
\Gamma = \frac{c^2 M}{32 \pi} \sqrt{1 - \frac{m_{\chi,\text{eff}}^2}{(M/2)^2}}\Theta\left(M^2-4m_{\chi,\text{eff}}^2\right),
\end{align}
where $m_{\chi,\text{eff}}^2 = \langle \lambda \tau^2 /2 \rangle$ is
the effective mass squared of $\chi$ and it is determined by the
time-dependent expectation value of $\tau$, as mentioned above.  Note
that the decay rate is non-zero only when the decay is kinematically
possible, i.e., $m_{\chi,\text{eff}} < M/2$, otherwise, it vanishes.
The critical value of triggeron field at which the decay channel opens
is $\tau_\text{c} = M/\sqrt{2\lambda}$.\footnote{To follow the evolutions of the mass of the daughter particles $\chi$
after the decay of $\phi$, we need to take into account the backreaction
of the particle production effect to the dynamics of $\tau$, whose
dedicated analysis is beyond the scope of this paper.
Once most of the energy density in $\phi$ has been converted to a
large number of $\chi$ particles when they are almost massless, energy
conservation implies that $\tau$ cannot move significantly.  A similar
backreaction effect is discussed in the context of
preheating~\cite{Kofman:2004yc}.  We expect that $\tau$ is trapped
around the origin and assume that the daughter particles $\chi$ remain
relativistic in the following analyses.
Even if the daughter particles do not behave as relativistic particles
due to its varying mass, the sudden reheating is realized in the case
where the daughter particles decay or annihilate to other light
particles, including the Standard Model particles, within a timescale
much shorter than the Hubble time at that time.
}

There are some conditions for this scenario to work.  Obviously, the
initial field value of triggeron $\tau_0$ should be large enough to
satisfy $\tau_0 > \tau_\text{c}$.  (We may assume $\tau_0 \geq 0$
without loss of generality.)  To make the reheating transition quick,
the speed of $\tau$ needs to be sufficiently large when it passes
through the critical point $\tau_\text{c}$, hence we assume
$\tau_0 \gg \tau_\text{c}$.  On the other hand, the triggeron field
should not dominate the energy density, and so $\tau_0$ should be much
less than the reduced Planck mass $M_{\text{P}}$. Thus, the required
condition for $\tau_0$ is
\begin{align}
\tau_\text{c}  \ll  \tau_0 \ll M_\text{P}.
\end{align}
Second, once the decay becomes kinematically possible, the typical
decay rate should be much larger than the Hubble scale,
$\Gamma \gg H$.  This requires $c^2 M \gg m$.  We also assume that
$\tau$ eventually decays into radiation before it would dominate the
energy density.

\begin{figure}
\centering
\includegraphics[width=0.4 \textwidth]{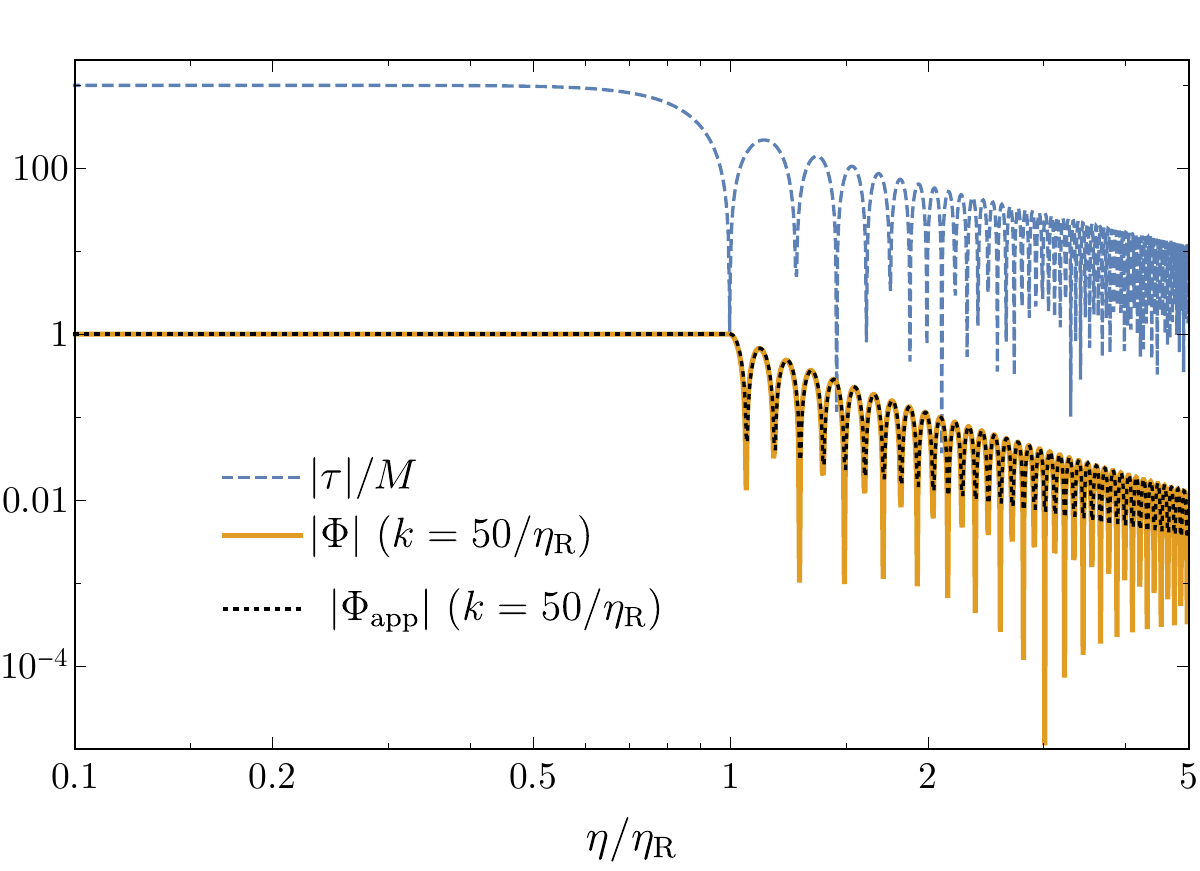}
\caption{Numerical results for the evolutions of the gravitational potential $\Phi$ and the triggeron $\tau$, normalized by $M$.
We also plot the analytical formula of $\Phi$, given by Eq.~(\ref{eq:phi_evo_eq}), with the black dotted line.
We take $\lambda = 0.1$, $c = 0.1$, and $\tau_0 = 1000 M$.
}
\label{fig:pertb_evo_inst_reh}
\end{figure}

Let us present the time evolution of the gravitational potential $\Phi$ to show how this model works. Figure~\ref{fig:pertb_evo_inst_reh} shows the evolution of $\Phi$ in addition to that of the field $\tau$. 
For $\Phi$, we use the equations for perturbations that are used in our accompanying paper~\cite{Inomata:2019zqy} to take into account the decay of $\phi$ to $\chi$.
This figure shows that the analytical expression of $\Phi$, given in Eq.~(\ref{eq:phi_formula}), is satisfactorily accurate in sudden-reheating scenarios.

%%%%%%%%%%%%%%%%%%
%%%%%%%%%%%%%%%%%%

Since $\phi$ and $\tau$ are independent degrees of freedom, fluctuations in
$\tau$ will introduce additional curvature perturbations and
non-Gaussianity due to the modulated reheating
mechanism~\cite{Kofman:2003nx,Dvali:2003ar,Zaldarriaga:2003my,Ackerman:2004kw,Podolsky:2005bw}.
To estimate those quantities, let us first note that the time
evolution of the triggeron is given by
$\tau = \tau_0 \sin (mt) / (mt)$.  The time when it reaches the
minimum ($\tau=0$) is $m t = \pi$, but it reaches the critical value
slightly before.  The decay time is thus estimated to be
\begin{align}
m t= & \pi \left( 1 -  \frac{\tau_{c}}{\tau_{0}} \right).
\end{align}
As discussed e.g. in Ref.~\cite{Kohri:2009ac}, the e-folding number is
related to the decay time as
\begin{align}
e^N  \propto t^{1/6}.
\end{align}
Thus, we can calculate $N' = (1/6) t'/t$ and
$N'' = (1/6) (t''/t - (t'/t)^2)$ where the prime denotes
differentiation with respect to $\tau_0$, and $t$ is evaluated at the
decay time. Explicitly,
\begin{align}
N' \simeq &  \frac{\tau_\text{c}}{6 \tau_0^2}, 
&  N''\simeq & - \frac{\tau_\text{c}}{3 \tau_0^3},
\end{align}
noting $\tau_{c}\ll \tau_0$. 
Using these values, we obtain
\begin{align}
\mathcal{P}_{\zeta^{(\tau)}} =& (N' \delta \tau_{0} )^2 \simeq  \frac{1}{36} 
  \left(  \frac{\tau_{c}}{\tau_{0}} \right)^2 \left( \frac{H_{\text{inf}}}{2 \pi \tau_{0}} \right)^2, \\
f_{\text{NL}} = & \frac{5}{6} \left( \frac{\mathcal{P}_{\zeta^{(\tau)}}}{\mathcal{P}_\zeta} \right)^2 \frac{N''}{(N')^2} \simeq - 10 
   \left( \frac{\mathcal{P}_{\zeta^{(\tau)}}}{\mathcal{P}_\zeta} \right)^2 \frac{\tau_{0}}{\tau_{c}},    
\end{align}
where $\zeta$ represents the total curvature perturbation, and
$\zeta^{(\tau)}$ is the contribution to $\zeta$ from $\tau$.  Note
that $f_{\mathrm{NL}}$ appears to contain a large factor
$\tau_{0}/\tau_{c}$, but the above expression implicitly contains the
inverse of this factor with a higher power. Hence, $f_{\text{NL}}$ can
be sufficiently small.  We conclude that non-Gaussianity can be small
enough to be consistent with observations provided that
$\tau_0 \gg \tau_\text{c}$ is satisfied.

Let us interpret the above model.  It is quite natural that the decay
of a field is prohibited by some symmetry.  For example, the lightest
particle charged under some unbroken symmetry is absolutely stable.  This is usually applied
to dark matter model building to explain its stability
\cite{Bertone:2004pz}. 
Thus, we assume that $\phi$ is charged under some symmetry.  If the
scalar field is real, as in the above toy model, the possible charge
assignment is limited, and so the scalar fields will be complex in a
more realistic situation.  In this context, $\tau$ must be assumed to
be a singlet (non-charged) with respect to the symmetry that protects
$\phi$'s stability because otherwise its initial nonzero expectation
value spontaneously breaks the symmetry.  $\chi$ can be interpreted as
some charged particle, initially heavier than $\phi$ due to its
$\tau$-dependent mass. However, it subsequently becomes lighter than
$\phi$, which triggers the $\phi$ decay.  For example, we can assign
$\phi$ charge $+2$ and $\chi$ charge $-1$.  Or, we can assign $\phi$
and $\chi$ the same charge and introduce a chargeless field $\chi'$
with an interaction such as $\phi \chi^{\dag} \chi'$.  In this way,
various generalizations of our simple model would be possible.  The
produced relativistic $\chi$ particles and antiparticles are assumed
to produce a thermal bath containing Standard Model particles through
scattering and annihilation, which reheats the Universe.

Alternatively, we may interpret $\tau$ as some symmetry breaking
field.  When a symmetry is broken, it is often the case that charged
fields (corresponding to $\chi$) become massive.  For example, the
Higgs mechanism makes gauge bosons massive.  In the Standard Model, it
also makes fermions massive through Yukawa interactions. One of the flat directions in the minimal supersymmetric standard model \cite{Enqvist:2003gh} would be a good candidate for
this purpose since most of the fields in the theory (corresponding to
$\chi$) can be massive when it obtains a finite expectation value.  In
this case, all the possible decay channels of $\phi$ must be
kinematically blocked or sufficiently suppressed.

\subsection{Another sudden-reheating scenario realized by a field that experiences a first order phase transition}

Suppose that $\phi$ is protected by a symmetry from decaying, without
any decay channels of $\phi$ to lighter particles. Let us further
assume that $\tau$ is charged under the symmetry and is too heavy for
$\phi$ to decay into.  There may be an interaction term of the form
\begin{align}
\mathcal{L} =  c \tau \phi \chi \chi + \dots,
\end{align}
where $c$ is a coupling constant. Suppose that initially the field
value of $\tau$ is zero, to be contrasted with the previous model.
Then the decay of $\phi$ becomes possible once $\tau$ acquires a
finite vacuum expectation value, thereby spontaneously breaking the
symmetry.

Such a symmetry-breaking phase transition can occur suddenly if the
phase transition is first order.  The transition occurs through the
tunneling effect, and the tunneling rate is exponentially sensitive to
the cosmic temperature (to be more precise, the temperature of the
thermal bath to which $\tau$ is coupled), and hence such a transition
is sudden \cite{Binetruy:2012ze}.  After the transition, $\phi$
becomes able to decay into $\chi$ particles.  Provided that this decay
rate is much larger than the Hubble parameter, the decay completes
within a timescale much shorter than the Hubble time at that time.
Associated with the decay of $\phi$, the temperature increases, which
may restore the symmetry temporarily.  Thus, the importance of the
backreaction to the decay of $\phi$ requires a further
study. Eventually, the temperature decreases, and $\tau$ settles to the
symmetry-breaking vacuum.

One way to suppress the backreaction may be to assume that the initial
thermal bath is made up of a hidden sector with $\tau$ being a portal
to the visible sector. Then, the increase in the temperature felt by
$\tau$ would not be significantly affected by the decay of $\phi$.

%%%%%%%%%%%%%%%%%%%%%%%%%%%%%%%%%%%%%%%%%%%%%%
\section{Approximate analytic formulas for induced GWs} \label{sec:analytic}

Here, we derive analytic formulas of the spectrum of induced GWs with some approximations based on sudden-reheating scenarios.  
During an eMD era, $\Phi$ is constant, and 
in the RD era, after reheating, the general solution of $\Phi$ is given by the sum involving spherical Bessel functions.  These solutions for the two epochs are connected at the transition.

As we can see in Fig.~\ref{fig:pertb_evo_inst_reh}, $\Phi$ can be well approximated by Eq.~(\ref{eq:phi_formula}) in sudden-reheating scenarios.  The explicit forms of the coefficients $A$ and $B$ in Eqs.~\eqref{eq:a_formula} and \eqref{eq:b_formula} are
\begin{align}
A(x_{\text{R}} ) =&  \left(- \frac{x_{\text{R}} ^2}{36} +1 \right)\cos \frac{x_{\text{R}} }{2\sqrt{3}}  + \frac{ \sqrt{3}}{6} x_{\text{R}}  \sin \frac{x_{\text{R}} }{2 \sqrt{3}} , \\
B(x_{\text{R}} ) = & \left( - \frac{x_{\text{R}} ^2}{36} +1 \right) \sin \frac{x_{\text{R}} }{2\sqrt{3}}  - \frac{\sqrt{3}}{6} x_{\text{R}}  \cos \frac{x_{\text{R}} }{2\sqrt{3}} .
\end{align}
Note that if we replace $x - x_{\text{R}} /2$ in the expression for
$\Phi$ by $x$, our formulas become those in the case of a pure RD era,
and hence we can use the general formulas in Appendix A of
Ref.~\cite{Kohri:2018awv}. We substitute the above $A$ and $B$ as well
as $C = - \cos (x-x_{\text{R}} /2)$ and $D= \sin (x-x_{\text{R}} /2)$
into these equations with $x_1$ and $x_2$ replaced by
$x_1 - x_{\text{R}} /2$ and $x_2 - x_{\text{R}} /2$, respectively.
The function $I$ is split into two terms as in
Eq.~\eqref{eq:i_formula_emd}.  The contributions generated during an
eMD era have been derived in Ref.~\cite{Assadullahi:2009nf} and
revised in Ref.~\cite{Kohri:2018awv}, and so we here mainly discuss the
contributions generated during the RD era, $I_{\text{RD}}$.  As
explained in the main text, this behaves very differently from the
counterpart for a pure RD era, which is obtained in the limit
$x_{\text{R}} \to 0$, because of fast oscillations of the modes that
are already inside the horizon at the reheating transition.
Extracting the redshift factor from the function,
$I_{\text{RD}} = \frac{1}{x - x_{\text{R}}/2}
\mathcal{I}_{\text{RD}}$,
we first calculate $\mathcal{I}_{\text{RD}}$, for which we can use the
results of Ref.~\cite{Kohri:2018awv}.

Below, we use two different approximations to obtain two main contributions.  The first approximation is valid for the large-scale modes with $k \ll k_{\text{max}}$, and the second approximation extracts the resonant contributions at $k \lesssim 2 k_{\text{max}}/\sqrt{3}$.  The sum of these two contributions turns out to explain the results of numerical integrations well.
For simplicity, we first consider the spectrum given in Eq.~\eqref{eq:pzeta_def} with $n_\text{s} = 1 $.  Generalization to cases with an arbitrary $n_\text{s}( >\!-3/2)$ is discussed at the end of this Appendix.

\subsection{Large-scale approximation}
As long as the scale $k^{-1}$ under consideration is much larger than the smallest scale $k_{\text{max}}^{-1}$, the integrations over $u$ and $v$, wavenumbers in units of $k$, are dominated by the large $t (\equiv u+v -1)$ region ($t \sim x_{\text{max,R}} / x_{\text{R}}$), hence $t x_{\text{R}}  \sim x_{\text{max,R}} \gg 1$.  
After taking late-time $(x\gg 1)$ oscillation average and changing variables from $u$ and $v$, to $t$ and $s \equiv u-v$,  we find 
\begin{widetext}
\begin{align}
\overline{\mathcal{I}_{\text{RD}}^2} \simeq &\frac{9 t^4 x_{\text{R}} ^8}{163840000} \left(  \pi^2 + \pi^2 \cos \frac{ s x_{\text{R}} }{\sqrt{3}} + 2 \text{Ci}\left( \frac{x_{\text{R}} }{2} - \frac{s x_{\text{R}} }{2\sqrt{3}} \right)^2 + 2  \text{Ci}\left( \frac{x_{\text{R}} }{2} + \frac{s x_{\text{R}} }{2\sqrt{3}} \right)^2 +  4 \cos \frac{ s x_{\text{R}} }{\sqrt{3}}  \text{Ci}\left( \frac{x_{\text{R}} }{2} + \frac{s x_{\text{R}} }{2\sqrt{3}} \right) \text{Ci}\left( \frac{x_{\text{R}} }{2} - \frac{s x_{\text{R}} }{2\sqrt{3}} \right) \right.  \nonumber \\
& + 2 \pi \sin \frac{ s x_{\text{R}} }{\sqrt{3}}  \left(   \text{Ci}\left( \frac{x_{\text{R}} }{2} + \frac{s x_{\text{R}} }{2\sqrt{3}} \right) -  \text{Ci}\left( \frac{x_{\text{R}} }{2} - \frac{s x_{\text{R}} }{2\sqrt{3}} \right) \right) - 2\pi    \left(1+\cos \frac{ s x_{\text{R}} }{\sqrt{3}} \right) \left(  \text{Si} \left( \frac{x_{\text{R}} }{2} - \frac{s x_{\text{R}} }{2\sqrt{3}} \right)  +  \text{Si} \left( \frac{x_{\text{R}} }{2} + \frac{s x_{\text{R}} }{2\sqrt{3}} \right) 
 \right)    \nonumber \\
& +4 \sin \frac{s x_{\text{R}} }{\sqrt{3}} \left( \text{Ci} \left( \frac{x_{\text{R}} }{2} - \frac{s x_{\text{R}} }{2\sqrt{3}} \right) \text{Si} \left( \frac{x_{\text{R}} }{2} + \frac{s x_{\text{R}} }{2\sqrt{3}} \right) -  \text{Ci} \left( \frac{x_{\text{R}} }{2} + \frac{s x_{\text{R}} }{2\sqrt{3}} \right) \text{Si} \left( \frac{x_{\text{R}} }{2} - \frac{s x_{\text{R}} }{2\sqrt{3}} \right)  \right) \nonumber \\
& \left. + 2 \text{Si} \left( \frac{x_{\text{R}} }{2} - \frac{s x_{\text{R}} }{2\sqrt{3}} \right)^2 + 2 \text{Si} \left( \frac{x_{\text{R}} }{2} + \frac{s x_{\text{R}} }{2\sqrt{3}} \right)^2  + 4 \cos \frac{ s x_{\text{R}} }{\sqrt{3}}  \text{Si}\left( \frac{x_{\text{R}} }{2} + \frac{s x_{\text{R}} }{2\sqrt{3}} \right) \text{Si}\left( \frac{x_{\text{R}} }{2} - \frac{s x_{\text{R}} }{2\sqrt{3}} \right)  \right) ,
 \label{curlyI_large-t}
\end{align}
\end{widetext}
where we have kept only terms with highest powers of $t$.
The sine and cosine integrals are defined as $\text{Si}(x) =
\int_0^x \text{d}z \sin (z) / z$ and $\text{Ci}(x) = - \int_x^\infty
\text{d} z \cos (z) / z$.
When we vary $s$, the above quantity varies approximately by a factor of two
at most.  However, the angular factor (the factor in the first line of
Eq.~\eqref{eq:p_h_formula}) in the large $t$ limit is $(s^2 - 1 )^2$,
which suppresses the nonzero $s$ part, and so it turns out that
setting $s=0$ is a good approximation for calculating
$\Omega_{\text{GW}}$ with 10\% errors at most.  If we set $s=0$, it is
simplified as
\begin{align}
\overline{\mathcal{I}_{\text{RD}}^2}|_{s=0} \simeq  & \frac{9 t^4 x_{\text{R}} ^8 \left( 4 \text{Ci}\left(\frac{x_{\text{R}} }{2}\right)^2 + \left( \pi - 2  \text{Si}\left(\frac{x_{\text{R}} }{2}\right) \right)^2 \right)}{81920000}  .\label{curlyI_large-t_s-0} 
\end{align}
This expression is so simple that we can analytically integrate it
over $t$ and $s$.  The integration region is $0 \leq s \leq 1$ and
$0 \leq t \leq -s+2 \frac{x_{\text{max,R}}}{x_{\text{R}}}-1$ for
$x_{\text{R}} \leq x_{\text{max,R}}$, and
$0 \leq s \leq 2 \frac{x_{\text{max, R}}}{x_{\text{R}} }-1$ and
$0 \leq t \leq -s+2 \frac{x_{\text{max,R}}}{x_{\text{R}} }-1$ for
$x_{\text{R}} > x_{\text{max,R}}$. For each case, there is also an
integration region obtained by the replacement $s\to -s$, but the
symmetry under this inversion ensures that the total result is
obtained by doubling the result obtained from the integration region
with $s>0$.  Then the GW spectrum under the large-scale (LS)
approximation is
\begin{widetext}
 \begin{align}
  \Omega_{\text{GW,RD}}^{\text{(LS)}} (\eta_c,k) \simeq &
\frac{ 4 \text{Ci}\left(\frac{x_{\text{R}} }{2}\right)^2 + \left( \pi - 2  \text{Si}\left(\frac{x_{\text{R}} }{2}\right) \right)^2 }{86016000000} A_\text{s}^2 x_{\text{R}} ^3 x_{\text{max,R}}^5   \times  \nonumber  \\ 
&
 \left( \Theta (x_{\text{max,R}}-x_{\text{R}} ) \left( 5376 - 17640 \widetilde{k} + 23760 \widetilde{k}^2 - 16425 \widetilde{k} ^3 + 5825 \widetilde{k}^4 - 847 \widetilde{k}^5  \right)    \right.   \nonumber \\
& \left.  \quad + \Theta (x_{\text{R}}  - x_{\text{max,R}}) \widetilde{k}^{-5} \left( 2   - \widetilde{k} \right)^6 \left( 4 - 8 \widetilde{k} - 9 \widetilde{k}^2 +13 \widetilde{k}^3 + 49 \widetilde{k}^4 \right) \right),
 \label{eq:Omega_GW_large-scale} 
 \end{align}
\end{widetext} 
where $\widetilde{k}=x_\text{R}/x_{\text{max,R}} = k/ k_{\text{max}}$.

\subsection{The resonant peak contributions}
It seems challenging to obtain a simple expression for the contributions from the $x_{\text{R}}  \simeq x_{\text{max,R}}$ region.
This is partly because the leading-order terms (with the highest power of $x_{\text{R}} $) cancel for generic values of $t$ and $s$, and the next-to-leading order terms are complicated.

Let us focus on a specific contribution that corresponds to the resonance-like peak (logarithmic divergence) at $t = \sqrt{3} - 1$ in the case of the monochromatic source in a pure RD era~\cite{Ananda:2006af}. 
The origin of the peak is the limit $x_{\text{R}}  \to 0$ of the Ci function.  In the present case, we do not take the limit $x_{\text{R}}  \to 0$, but instead we focus on contributions from the region where the integration variable $t$ hits the logarithmic singularity of the function Ci, possibly causing an enhancement.
For this purpose, we do not take the large $t$ limit.  Instead, we can take the large $x_{\text{R}}$ limit since it turns out that this effect is most efficient for the smallest-scale modes.

We focus on the terms containing the Ci function whose argument can vanish, neglecting the other terms.
Furthermore, we take the late time limit $x \to \infty$ as well as oscillation average.
With these approximations, we find
\begin{align}
\overline{\mathcal{I}_{\text{RD}}^2} \approx Y\frac{9 (-5+s^2+2t+t^2)^4 x_{\text{R}} ^8}{81920000 (1-s+t)^2 (1+s+t)^2}  \text{Ci}\left( |y| \right)^2 ,
\end{align} 
where  $y \equiv (t-\sqrt{3}+1)x_{\text{R}} /(2\sqrt{3})$ and $Y$ is a fudge factor of order unity. We focus on spiky contributions around $y =0$ or equivalently $t = \sqrt{3}-1$.  Except for the argument of the Ci function, we may set $t = \sqrt{3}-1$, which enables us to do the integration over $s$.
Then, the resonant contribution to the GW spectrum is
\begin{align}
\Omega_{\text{GW,RD}}^{\text{(res)}} (\eta_c,k) \simeq & 
Y\int _{-s_0 (x_\text{R})}^{s_0 (x_\text{R})} \text{d} s \frac{3(1-s^2)^2 }{81920000} A_{\text{s}}^2 x_{\text{R}} ^8 \nonumber \\
& \times 2  \int_{0}^1 \text{d} y \text{Ci}(y)^2 \frac{2\sqrt{3}}{x_{\text{R}} }  \nonumber \\
=& Y\frac{ 2.30285 }{102400000} \sqrt{3} A_{\text{s}}^2 x_{\text{R}} ^7  s_0 (x_\text{R})  \nonumber \\
& \times \left( 15-10 s_0^2 (x_\text{R}) + 3 s_0^4  (x_\text{R})\right)   ,\label{eq:Omega_GW_resonance}
\end{align}
where
\begin{align}
s_0 (x_\text{R}) =& \begin{cases}
1  &   x_\text{R} \leq \frac{2 x_\text{max,R}}{1+\sqrt{3}} \\
2 \frac{x_\text{max,R}}{x_\text{R}} -\sqrt{3} & \frac{2 x_\text{max,R}}{1+\sqrt{3}} \leq x_\text{R} \leq \frac{2 x_\text{max,R}}{\sqrt{3}} \\
0 &  \frac{2 x_\text{max,R}}{\sqrt{3}} \leq x_\text{R} 
\end{cases}. \label{eq:s_0}
\end{align}
In the first equality in Eq.~\eqref{eq:Omega_GW_resonance}, we have changed the integration variable from $t$ to $y$ with the Jacobian factor $2\sqrt{3}/x_{\text{R}} $.
This integration is for the spiky part, and so we limit the integration region to $|y|<1$. 
The choice of the integration boundary here is somewhat arbitrary, but this uncertainty can also be absorbed by the fudge factor $Y$. 
We determine the value of $Y$ by comparing Eq.~(B7) with the numerical result.  
We find that $Y=2.3$ is a good fit, so we set $Y = 2.3$ throughout this work. 

\begin{figure}[t]
 \centering
{\includegraphics[width=0.4 \textwidth]{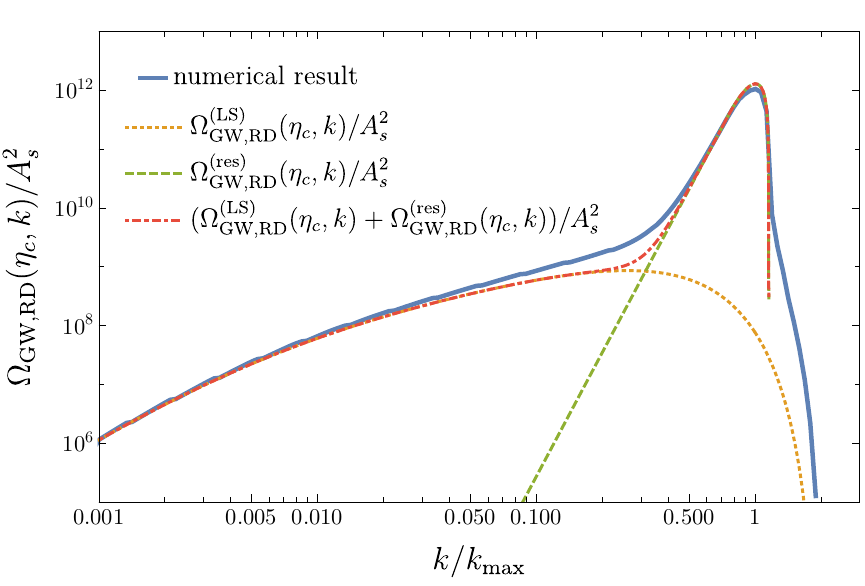}}
  \caption{Comparison of the analytic and numerical results for the induced GWs.  The blue solid line shows the numerical result. The orange dotted, greed dashed, and red dot-dashed lines show the large-scale approximation [Eq.~\eqref{eq:Omega_GW_large-scale}], the resonant contribution [Eq.~\eqref{eq:Omega_GW_resonance}], and their sum, respectively.
  We take the power spectrum given in Eq.~(\ref{eq:pzeta_def}) with $k_\text{max}=450/\eta_\R$ and $n_s=1$.
  }
 \label{fig:Omega_GW_MD2RD_cutoff3}
  \end{figure}

The total spectrum is approximated by the contribution produced after the reheating transition, $\Omega_\text{GW} \simeq \Omega_\text{GW,RD}$, which is given by the sum of Eqs.~\eqref{eq:Omega_GW_large-scale} and \eqref{eq:Omega_GW_resonance}:
\begin{align}
\label{eq:omega_gw_approximation}
\Omega_{\text{GW,RD}} \simeq \Omega_\text{GW,RD}^\text{(LS)} +  \Omega_\text{GW,RD}^\text{(res)} .
\end{align}
This is compared with the numerical result in Fig.~\ref{fig:Omega_GW_MD2RD_cutoff3}.
From this figure, we can see that those approximate analytic formulas fit the numerical result very well.

The $k$ dependence of $\Omega_{\text{GW}}$ is summarized as follows.
It is proportional to $k^3$, neglecting a logarithmic factor, for $k \lesssim 1/\eta_{\text{R}}$, then it scales as $k$ for $k \gtrsim 1/\eta_\R $.  The slope of the resonant contribution is $k^7$, which peaks at $k \simeq 2 k_{\text{max}}/\sqrt{3}$. Finally, it decrease sharply, and vanishes at $k = 2 k_{\text{max}}$. This behavior is summarized in Eq.~\eqref{eq:Omega_GW_rough_behavior}.

\vspace{25pt}
  \subsection{Approximate analytic formulas for induced GWs from power-law primordial spectra} \label{sec:analytic2}
We can generalize our calculations to power-law primordial spectra with a cutoff [see Eq.~(\ref{eq:pzeta_def})]. 
We can use the formulas of $\overline{\mathcal{I}_\text{RD}^2}$ obtained above.  %, but we need to include the factor
Using the large $t$ approximation and assuming $n_\text{s} > - 3/2$, we obtain the following expression:
\begin{widetext}
\begin{align}
\label{eq:omega_tilt_ls}
\Omega_{\text{GW,RD}}^{\text{(LS)}}\simeq & \frac{3 \left( 4 \text{Ci}\left(\frac{x_R}{2}\right)^2 + \left( \pi - 2  \text{Si}\left(\frac{x_R}{2}\right) \right)^2 \right) A_\text{s}^2  x_\text{max,R}^8 }{2^{17+2 n_\text{s}} \times 625 (3+2n_\text{s})} \left( \frac{2 x_{\text{max,R}}}{x_\text{R}}-1 \right)^{2 n_\text{s}} \left( \frac{x_\text{R}}{x_{*,\text{R}}} \right)^{2(n_s -1)} \nonumber \\
& \qquad \times \left( \widetilde\Omega_{\text{GW,RD}}^{\text{(LS,1)}} \Theta (x_{\text{max,R}} - x_\text{R}) + \widetilde\Omega_{\text{GW,RD}}^{\text{(LS,2)}} \Theta (x_\text{R} - x_{\text{max,R}} )  \right)  \Theta ( 2 x_{\text{max,R}} - x_\text{R}),
\end{align}
where 
\begin{align}
\label{eq:omega_tilt_ls1}
\widetilde\Omega_\text{GW,RD}^\text{(LS,1)} =& \frac{1}{(2+n_\text{s}) (3 + n_\text{s}) (4 + n_\text{s}) (5 + 2 n_\text{s}) (7 + 2 n_\text{s})}\times \left( 1536 - 6144 \widetilde{k} + (7168-1920 n_\text{s} -256 n_\text{s}^2 ) \widetilde{k}^2 \phantom{\left(\frac{\widetilde{k}}{\widetilde{k}}\right)^{n_s}} \right. \nonumber \\
&  +(5760 n_\text{s} + 768 n_\text{s}^2) \widetilde{k}^3 +(1328 n_\text{s} + 3056 n_\text{s}^2 + 832 n_\text{s}^3 +64 n_\text{s}^4) \widetilde{k}^4  \nonumber \\
& - (7168+12256 n_\text{s} + 7392 n_\text{s}^2 +1664 n_\text{s}^3 + 128 n_\text{s}^4 ) \widetilde{k}^5 +(7392 + 10992 n_\text{s} + 5784 n_\text{s}^2 + 1248 n_\text{s}^3 + 96 n_\text{s}^4) \widetilde{k}^6   \nonumber   \\
&  -(2784+ 3904 n_\text{s} + 1960 n_\text{s}^2 + 416 n_\text{s}^3 + 32 n_\text{s}^4 ) \widetilde{k}^7 + (370+503 n_\text{s} + 247 n_\text{s}^2 +52 n_\text{s}^3 + 4n_\text{s}^4) \widetilde{k}^8 \nonumber \\
& \left. - 256 ( 1 - \widetilde{k})^6 (6 + 6 (2 +n_\text{s}) \widetilde{k} + (2 + n_\text{s})(5+2 n_\text{s}) \widetilde{k}^2 ) \left( 1 - \frac{ \widetilde{k}}{2 - \widetilde{k}} \right)^{2 n_\text{s}} \right),  \\
\label{eq:omega_tilt_ls2}
\widetilde\Omega_{\text{GW,RD}}^{\text{(LS,2)}}=& 2(2 -\widetilde{k})^4 \Gamma (4+2 n_\text{s}) \left( \frac{\widetilde{k}^4}{\Gamma(5+2n_\text{s})} -\frac{4 \widetilde{k}^2 (2 -\widetilde{k})^2}{\Gamma(7+2n_\text{s})} + \frac{24(2 -\widetilde{k})^4}{\Gamma(9+2 n_\text{s})} \right),
\end{align}
with $\Gamma (x)$ denoting the Gamma function.
The other important component, the resonance contribution, is obtained as
\begin{align}
\label{eq:omega_tilt_res}
\Omega_\text{GW,RD}^\text{(res)} =& Y\frac{2.30285 \times \sqrt{3} \, 3^{n_\text{s}}}{2^{13+2 n_\text{s}} \times 625 } x_\text{R}^7 \left( \frac{x_\text{R}}{x_\text{*,R}} \right)^{2(n_\text{s}-1)} s_0(x_\text{R}) \nonumber \\
& \times \left( 4 {}_2 F_1 \left( \frac{1}{2}, 1- n_\text{s} ; \frac{3}{2} ; \frac{s_0^2 (x_\text{R})}{3} \right)   -3 {}_2 F_1 \left( \frac{1}{2}, - n_\text{s} ; \frac{3}{2} ; \frac{s_0^2 (x_\text{R})}{3} \right) - s_0^2 (x_\text{R}) {}_2 F_1 \left( \frac{3}{2}, - n_\text{s} ; \frac{5}{2} ; \frac{s_0^2 (x_\text{R})}{3} \right)   \right),
\end{align}
where ${}_2 F_1 \left(a,b; c; z\right) $ is the hypergeometric
function, and $s_0 (x_\text{R})$ is defined in Eq.~\eqref{eq:s_0}.
The total spectrum is again approximated by the sum:
$\Omega_{\text{GW}} \simeq \Omega_\text{GW,RD}^\text{(LS)} +
\Omega_\text{GW,RD}^\text{(res)} $.

For completeness and comparison, we also present the formulas of the component of the induced GWs produced during the eMD era, $\Omega_\text{GW,eMD}$.  Such a formula is presented in Ref.~\cite{Kohri:2018awv} for the scale-invariant power spectrum of the curvature perturbations with a cutoff scale $k_{\text{max}}$.  We generalize it to the power-law spectrum with a cutoff scale.
For this purpose, we use the large $t$ approximation once again.
\begin{align}
\Omega_{\text{GW,eMD}}^\text{(LS)}(\eta_c, k)=&   \frac{3 R(x_\text{R})   A_\text{s}^2 \left( \frac{x_\text{R}}{x_{\text{*,R}}} \right)^{2(n_\text{s}-1)}  \Theta ( 2 x_{\text{max,R}} - x_\text{R}) }{25 \times 2^{1+2n_\text{s}} n_\text{s} (1+n_\text{s})(2+n_\text{s})(-1+2n_\text{s})(1+2n_\text{s})(3+2n_\text{s}) \widetilde{k}^{4+2n_\text{s}} }  \times \nonumber \\
&  \left( -4^{2+ n_\text{s}} \left(1-\widetilde{k} \right)^{2(1+n_\text{s})} \left( 6 + 6 n_{\text{s}} \widetilde{k}+ n_\text{s} (1+2 n_\text{s} )\widetilde{k}^2 \right)  \Theta (1-\widetilde{k})  \right.  \nonumber \\
& \left. + \left(2 - \widetilde{k}\right)^{2 n_\text{s}} \left( 96 -192 \widetilde{k} + (96-8 n_\text{s} (7+2 n_\text{s})) \widetilde{k}^2 + 8 n_\text{s}  (7+2 n_\text{s})  \widetilde{k}^3 + n_\text{s} (1+2 n_\text{s}) (11 + n_\text{s} (9+2 n_\text{s}))   \widetilde{k}^4   \right)  \right),
\end{align}
\end{widetext}
where $R(x_\text{R})$ is the relative suppression factor given by Eq.~(45) of Ref.~\cite{Kohri:2018awv}.  In the scale-invariant case ($n_\text{s}=1$), this reproduces the leading term  in the large-scale limit $k \ll k_\text{max}$ of the formula in Ref.~\cite{Kohri:2018awv} up to the factor $1/4$ revised in our accompanying paper~\cite{Inomata:2019zqy}. 
The above formula is derived assuming $n_\text{s} > 1/2$.
There are no resonance contributions for $\Omega_{\text{GW,eMD}}$, so that the total spectrum can be approximated by the above formula, $\Omega_{\text{GW,eMD}}\simeq \Omega_{\text{GW,eMD}}^\text{(LS)}$ for $k \ll k_\text{max}$.
However, this contribution is subdominant compared to $\Omega_\text{GW,RD}$ as shown in Fig.~\ref{fig:emd_gw_profiles} for $n_\text{s}=1$.

%%%%%%%%%%%%%%%%%%%%%%%%%%%%%%%%%%%
%%%%%%%%%%%%%%%%%%%%%%%%%%%%%%%%%%%
%%%%%%%%%%%%%%%%%%%%%%%%%%%%%%%%%%%
\small
\bibliographystyle{apsrev4-1}
\bibliography{gw_emd_enh}

\end{document}